\newif\ifpreprint

\preprinttrue 

\ifpreprint
\documentclass[journal=jctcce,manuscript=article]{achemso}
\else
\documentclass[journal=jctcce,manuscript=article,layout=twocolumn]{achemso}
\fi

\usepackage{amsmath}
\usepackage{amssymb}
\usepackage{enumitem}

\usepackage{newtxtext,newtxmath}

\usepackage{braket}
\usepackage{ulem}
\usepackage{threeparttable}
\usepackage{multirow}
\usepackage[version=4]{mhchem}
\usepackage{xcolor}
\usepackage{xspace}
\usepackage{ifthen}
\usepackage{indentfirst}
\usepackage{titlesec}
\usepackage[colorlinks = true,
            linkcolor = blue,
            urlcolor  = black,
            citecolor = blue,
            anchorcolor = black]{hyperref}

\definecolor{goodorange}{RGB}{225,125,0}
\definecolor{goodgreen}{RGB}{5,130,5}
\definecolor{goodred}{RGB}{220,50,25}
\definecolor{goodblue}{RGB}{30,144,255}

\usepackage{dcolumn}
\newcolumntype{d}[1]{D{.}{.}{#1}}
\makeatletter
\newcolumntype{B}[3]{>{\boldmath\DC@{#1}{#2}{#3}}c<{\DC@end}}
\makeatother

\newcommand*{\sunit}{$E_{\rm h}^{-2}$\xspace}
\newcommand*{\Eh}{$E_{\rm h}$\xspace}

\newcommand*{\PSI}{{\scshape Psi4}\xspace}

\newcommand*{\forte}{{\scshape Forte}\xspace}

\newcommand{\tens}[3]{{#1}_{#2}^{#3}}
\newcommand{\dfock}[1]{\epsilon_{#1}}
\newcommand{\cop}[1]{\hat{a}^\dag_{#1}}
\newcommand{\aop}[1]{\hat{a}_{#1}}
\newcommand{\sqop}[2]{\hat{a}_{#2}^{#1}}

\newcommand{\kro}[2]{\delta_{#2}^{#1}}

\newcommand{\no}[1]{ \{ {#1} \}}

\newcommand{\squop}[2]{\hat{E}^{#1}_{#2}}  

\newcommand{\dadapt}[2]{\Gamma_{#2}^{#1}}
\newcommand{\cdadapt}[2]{\Theta_{#2}^{#1}}
\newcommand{\cuadapt}[2]{\Lambda_{#2}^{#1}}

\makeatletter

\newbox\swb@xone
\newbox\swb@xtwo
\newbox\swb@xthree
\newbox\swb@xfour
\newdimen\swdimen@ne
\newdimen\swdimentw@

\newcommand{\acontraction}[5][1ex]{%
  \mathchoice
    {\acontraction@\displaystyle{#2}{#3}{#4}{#5}{#1}}%
    {\acontraction@\textstyle{#2}{#3}{#4}{#5}{#1}}%
    {\acontraction@\scriptstyle{#2}{#3}{#4}{#5}{#1}}%
    {\acontraction@\scriptscriptstyle{#2}{#3}{#4}{#5}{#1}}}%
\newcommand{\acontraction@}[6]{%
  \setbox\swb@xone=\hbox{${}#1{}#2{}$}%
  \setbox\swb@xtwo=\hbox{${}#1{}#3{}$}%
  \setbox\swb@xthree=\hbox{${}#1{}#4{}$}%
  \setbox\swb@xfour=\hbox{${}#1{}#5{}$}%
  \swdimen@ne=\wd\swb@xtwo%
  \advance\swdimen@ne by \wd\swb@xfour%
  \divide\swdimen@ne by 2%
  \advance\swdimen@ne by \wd\swb@xthree%
  \vbox{%
    \hbox to 0pt{%
      \kern \wd\swb@xone%
      \kern 0.5\wd\swb@xtwo%
      \acontraction@@{\swdimen@ne}{#6}%
      \hss}%
    \vskip 0.5ex
    \vskip\ht\swb@xtwo}}

\newcommand{\acontraction@@}[3][0.05em]{%
  \hbox{%
    \vrule width #1 height 0pt depth #3%
    \vrule width #2 height 0pt depth #1%
    \vrule width #1 height 0pt depth #3%
    \relax}}

\newcommand{\tcontraction}[5][1ex]{%
  \mathchoice
    {\tcontraction@\displaystyle{#2}{#3}{#4}{#5}{#1}}%
    {\tcontraction@\textstyle{#2}{#3}{#4}{#5}{#1}}%
    {\tcontraction@\scriptstyle{#2}{#3}{#4}{#5}{#1}}%
    {\tcontraction@\scriptscriptstyle{#2}{#3}{#4}{#5}{#1}}}%
\newcommand{\tcontraction@}[6]{%
  \setbox\swb@xone=\hbox{${}#1{}#2{}$}%
  \setbox\swb@xtwo=\hbox{${}#1{}#3{}$}%
  \setbox\swb@xthree=\hbox{${}#1{}#4{}$}%
  \setbox\swb@xfour=\hbox{${}#1{}#5{}$}%
  \swdimen@ne=\wd\swb@xtwo%
  \advance\swdimen@ne by \wd\swb@xfour%
  \divide\swdimen@ne by 2%
  \advance\swdimen@ne by \wd\swb@xthree%
  \vbox{%
    \hbox to 0pt{%
      \kern \wd\swb@xone%
      \kern 0.5\wd\swb@xtwo%
      \tcontraction@@{\swdimen@ne}{#6}%
      \hss}%
    \vskip 0.5ex
    \vskip\ht\swb@xtwo}}

\newcommand{\tcontraction@@}[3][0.075em]{%
  \hbox{%
    \vrule width #1 height 0pt depth #3%
    \vrule width #2 height 0pt depth #1%
    \vrule width #1 height 0pt depth #3%
    \relax}}

\makeatother

\usepackage[fontsize=11pt]{scrextend}
\author{Chenyang Li}
\email{chenyang.li@bnu.edu.cn}
\affiliation{Key Laboratory of Theoretical and Computational Photochemistry, Ministry of Education, College of Chemistry, Beijing Normal University, Beijing 100875, China}

\author{Shuxian Mao}
\affiliation{Key Laboratory of Theoretical and Computational Photochemistry, Ministry of Education, College of Chemistry, Beijing Normal University, Beijing 100875, China}

\author{Renke Huang}
\affiliation{Department of Chemistry and Cherry Emerson Center for Scientific Computation, Emory University, Atlanta, GA 30322, USA}

\author{Francesco A. Evangelista}
\affiliation{Department of Chemistry and Cherry Emerson Center for Scientific Computation, Emory University, Atlanta, GA 30322, USA}

\let\oldmaketitle\maketitle
\let\maketitle\relax

\title{Frozen natural orbitals for the state-averaged driven similarity renormalization group}

\begin{document}

\ifpreprint
\else
\twocolumn[
\begin{@twocolumnfalse}
\fi
\oldmaketitle

\begin{abstract}
We present a reduced-cost implementation of the state-averaged driven similarity renormalization group (SA-DSRG) based on the frozen natural orbital (FNO) approach.
The natural orbitals (NOs) are obtained by diagonalizing the one-body reduced density matrix from SA-DSRG second-order perturbation theory (SA-DSRG-PT2).
Subsequently, the virtual NOs with occupation numbers (ONs) smaller than a user-defined threshold are excluded from the high-level electron correlation treatment beyond SA-DSRG-PT2.
An additive second-order correction is applied to the SA-DSRG Hamiltonian to reintroduce the correlation effect from the discarded orbitals.
The FNO SA-DSRG method is benchmarked on 35 small organic molecules in the QUEST database.
When keeping 98--99\% of the cumulative ONs, the mean absolute error in the vertical transition energies due to FNO is less than 0.01~eV.
Using the same FNO threshold, we observe a speedup of 9 times the conventional SA-DSRG implementation for nickel carbonyl with a quadruple-$\zeta$ basis set.
The FNO approach enables third-order SA-DSRG perturbation theory computations on chloroiron corrole \ce{[FeCl(C19H11N4)]} with more than 1000 basis functions.
\end{abstract}

\ifpreprint
\else
\end{@twocolumnfalse}
]
\fi


\ifpreprint
\else
\small
\fi

\noindent

\section{Introduction}
\label{sec:intro}

One important problem in quantum chemistry is accurately characterizing electronically excited states, as it necessitates careful modeling of electron correlation and orbital relaxation effects.
A variety of excited-state methods have been developed.
Among the most efficient and widely used approaches include $\Delta$ self-consistent field (SCF)\cite{Jones1989,Gilbert:2008fz} and time-dependent density functional theory.\cite{Runge1984,Casida2012,Maitra2016}
More involved methods incorporate electron correlation effects in a systematically improvable way, such as excited-state-specific methods,\cite{Hardikar:2020hf,Clune:2020tn,Tuckman2023} algebraic diagrammatic construction (ADC),\cite{Schirmer:1982cf,Mertins:1996eu,Dreuw:2014hd,Dreuw2023} and linear response (LR) and equation-of-motion (EOM) formulations of coupled cluster (CC) theory.\cite{Monkhorst:1977gs,Sekino:1984fm,Koch:1990du,Rowe:1968el,Stanton:1993be,Krylov:2008bh}
However, the accuracy of these single-reference methods is often compromised when the ground state has significant multireference (MR) character or the excited state is dominated by multi-electron excitations.\cite{Loos2019}
Genuine MR methods provide a more straightforward yet general approach for excited states by starting from a multideterminantal zeroth-order state that can properly deal with these situations.

In the MR framework, one normally starts from a zeroth-order multideterminantal reference wave function where the set of molecular orbitals (MOs) are optimized via the multi-configurational SCF (MCSCF) procedure.\cite{Docken1972}
The configuration space of the reference wave function, termed active space, can be constructed in many ways.
The most common one is the complete active space (CAS),\cite{Cheung1979,Roos:1980fd} which contains all possible configurations by distributing a small number of electrons into a set of selected orbitals (i.e., active orbitals).
On top of this reference wave function, electron correlation effects are introduced via configuration interaction,\cite{Szalay:2012df,Werner:1988ku,Sivalingam:2016hr} perturbation theory (MRPT),\cite{Andersson:1992cq,Angeli:2001bg,Hoffmann:2009ih,Mazin2021} or CC (MRCC).\cite{Lyakh:2012cn,Koehn:2013cp,Evangelista:2018bt}
However, these MR generalizations suffer from a few technical difficulties.
Examples include the intruder-state problem\cite{Paldus:1993dx,Roos:1995jz} and the multiple parentage issue\cite{Evangelista:2018bt} in various MRPTs and MRCC approaches, resulting in non-converging equations and discontinuous potential energy surfaces.\cite{Kowalski:2000cj}
Another crucial aspect lies in the high numerical complexity of the working equations.
For instance, the internally contracted MRCC with singles and doubles involves more than one million terms in the amplitude equations,\cite{Hanauer:2011ey} while there are only 45 terms for the single-reference counterpart.

The steep cost of post-MCSCF methods is rooted in the exponential growth of the active space and the treatment of electron correlation.
The former can be alleviated by the density matrix renormalization group\cite{White:1993fb,Chan:2011gb} and various selected configuration interaction schemes.\cite{Evangelisti1983,Schriber:2016kl,Holmes:2016fm,Tubman2020}
For the latter, many techniques developed for reducing the cost of single-reference correlation methods can be introduced into the MR regime.
One effective approach is to factorize the two-electron integrals into contractions of low-rank tensors,\cite{Whitten:1973ju,Dunlap:1979gh,Koch:2003go,Aquilante:2009eu,Zhang2021c,Hohenstein2012b,*Parrish2012,*Hohenstein2012a,*Parrish:2013iz} which enables second-order MRPT (MRPT2) computations to be performed routinely on molecules with 2000 basis functions.\cite{Aquilante:2008gk,Bostrom:2010is,Hannon:2016bh,Song:2018fi,Song:2020jf}
Even larger systems can be simulated using MRPT2s when combining them with local correlation methods based on pair natural orbitals (PNOs).\cite{Guo:2016eh,Menezes:2016gu,Kats:2019bs,Saitow2022}
The PNO formalism generates a compact correlation space for each electron pair individually, and the computational cost is reduced by truncating the virtual orbital space according to the natural occupancy.

A simpler approach to reduce the size of virtual orbitals is given by the global frozen natural orbitals (FNOs).\cite{Barr1970,Sosa1989}
In this method, the one-particle reduced density matrix (1-RDM) of a low-level correlated wave function is diagonalized and orbitals with small natural occupation numbers are discarded for subsequent high-level electron correlation computations.
Successful applications of FNO are mostly found in the CC hierarchy,\cite{Taube:2005hv,DePrinceIII:2013ki,Deprince2013,Gyevi-Nagy2021,Nagy2021} where 20--60\% of the virtual orbitals can be removed without sacrificing the accuarcy significantly for a triple-$\zeta$ basis set.\cite{Taube:2005hv}
More recently, FNO-based excited-state methods have also been formulated, including LR/EOM-CC,\cite{Landau2010,Pokhilko:2020cb,Mester2017} ADC,\cite{Mester2018,Mukhopadhyay2023} and $GW$ theories.\cite{Monzel2023}
For MR methods, the FNO scheme has not been widely applied with the notable exception of second-order CAS perturbation theory (CASPT2)\cite{Aquilante:2009ej,Segarra-Marti2015} and MR-EOM-CC.\cite{Huntington:2015fq}
Nevertheless, these FNO-MR methods approximate the 1-RDM with the one from second-order M{\o}ller--Plesset perturbation theory.

In this work, we construct FNOs using the 1-RDM from a genuine MRPT2, derived from the driven similarity renormalization group (DSRG).\cite{Evangelista:2014kt,Li:2019fu,Li:2015iz}
Then, the virtual orbital space is truncated to reduce the computational cost of DSRG third-order MRPT (MRPT3)\cite{Li:2017bx} and the linearized DSRG with singles and doubles [LDSRG(2)],\cite{Li:2016hb} both of which scale as the fourth power to the number of virtual orbitals.
Recent benchmarks show that DSRG-MRPT3 and MR-LDSRG(2) yield comparable accuracy to EOM-CC with singles and doubles for valence excitations,\cite{Wang2023} while they excel in predicting core-excited states.\cite{Huang2022,Huang2022d}
In the following, we shall focus on the state-averaged (SA) variant of DSRG,\cite{Li:2018kl} which assumes that all targeted states share similar electron correlation effects.
Unlike the state-specific formalism, SA-DSRG can describe conical intersections correctly.

This paper is organized as follows.
In Sec.~\ref{sec:theory}, we briefly review the SA-DSRG ansatz and discuss its FNO formulation.
To illustrate the capability of the FNO-based SA-DSRG, vertical transition energies of small organic molecules (Sec.~\ref{subsec:questdb}) and transition-metal complexes (Sec.~\ref{subsec:tm}) are presented.
Finally, we conclude and discuss future developments in Sec.~\ref{sec:conclusion}.

\section{Theory}
\label{sec:theory}

We first summarize some key aspects of the SA-DSRG formalism (see Ref.~\citenum{Li:2018kl} for details).
In SA-DSRG, we start from an ensemble of $n$ reference electronic states $\{\Psi_\alpha | \alpha = 1, 2, \dots, n\}$.
To describe static correlation effects, these states are obtained via an SA complete-active-space self-consistent field (CASSCF) computation.
The SA-DSRG ansatz then considers the remaining dynamical correlation effects by transforming the bare Hamiltonian $\hat{H}$ via a unitary transformation [$\hat{U}(s)$]:
\begin{align}
\label{eq:Hbar_trans}
\hat{H} \rightarrow \bar{H}(s) = \hat{U}^{\dag} (s) \hat{H} \hat{U} (s), \quad s \geq 0,
\end{align}
where $s \in [0, \infty)$ is the so-called flow parameter.
As $s$ increases, the DSRG transformation gradually reduces the magnitude of the nondiagonal components of $\bar{H}(s)$ [$\bar{H}^{N}(s)$] to zero.
For finite values of $s$, the DSRG Hamiltonian possesses a band-diagonal structure in the Fock space.\cite{Evangelista:2014kt,Li:2019fu}
The unitary transformation $\hat{U}(s)$ is parametrized using the $s$-dependent cluster operator $\hat{T}(s)$:
\begin{align}
\hat{U}(s) = \exp[\hat{A}(s)], \quad \hat{A}(s) = \hat{T}(s) - \hat{T}^\dag(s) .
\end{align}
Then, the DSRG Hamiltonian $\bar{H}(s)$ can be expanded by employing the Baker--Campbell-Hausdorff (BCH) formula, with each commutator evaluated using the generalized Wick's theorem of Mukherjee and Kutzelnigg.\cite{Mukherjee:1997tk,Kutzelnigg:1997ut}
The DSRG cluster amplitudes $\hat{T}(s)$ are determined via a set of many-body conditions\cite{Li:2016hb,Datta:2011ca}
\begin{align}
\label{eq:dsrg_flow}
\bar{H}^{N}(s) = \hat{R} (s),
\end{align}
where the source operator $\hat{R} (s)$ specifies the evolution of the many-body components of $\bar{H}^{N}(s)$ as the flow parameter grows.
Finally, $\bar{H}(s)$ is diagonalized in the complete active space to obtain the SA-DSRG energies of the targeted states:
\begin{align}
\label{eq:Hbar_relax}
\bar{H}(s) \ket{\Psi'_\alpha (s)} = E'_\alpha \ket{\Psi'_\alpha (s)}.
\end{align}
Equation \eqref{eq:Hbar_relax} generates a set of relaxed states, which can be taken as the reference ensemble for SA-DSRG to solve for a new set of cluster amplitudes.
This reference relaxation procedure may be iterated till convergence, but energies accurate to one milliHartree are often obtained by solving Eqs.~\eqref{eq:dsrg_flow} and \eqref{eq:Hbar_relax} twice.\cite{Li:2017bx}
We refer to this doubly relaxed version as the relaxed SA-DSRG hereafter.

Equations \eqref{eq:Hbar_trans}--\eqref{eq:Hbar_relax} formally define the SA-DSRG formalism, yet approximations must be introduced for practical applications.
In the linearized DSRG with one- and two-body operators [LDSRG(2)] approach,\cite{Li:2016hb} the cluster operator is truncated to singles and doubles and the BCH expansion is computed via the recursive linear commutator approximation:\cite{Yanai:2006gi}
\begin{align}
\label{eq:bch12}
\bar{H}(s) &\approx \bar{H}_{0,1,2} (s) \notag\\
&= \hat{H} + \sum_{k=1}^{\infty} \frac{1}{k!} \underbrace{[\cdots[[\hat{H}, \hat{A}(s)]_{0,1,2}, \hat{A}(s)]_{0,1,2}, \cdots]_{0,1,2}}_{k\text{-nested commutators}}.
\end{align}
In this equation, the subscript ``$0,1,2$'' indicates that the preceding commutator includes the one- and two-body operators as well as the fully contracted scalar term.
The nested commutators are terminated such that the Frobenius norm of the last nested commutator is numerically insignificant (e.g., $< 10^{-12}$).
The computational cost of evaluating $\bar{H}_{0,1,2}(s)$ is dominated by the contractions between a two-body intermediate operator $\hat{O}_2 (s)$ and the doubles cluster operator $\hat{T}_2(s)$, yielding another two-body intermediate $\hat{C}_2(s)$:
\begin{align}
\label{eq:O2T2C2}
\hat{C}_2(s) = [\hat{O}_2 (s), \hat{T}_2(s)]_2.
\end{align}
Computing Eq.~\eqref{eq:O2T2C2} scales asymptotically as ${\cal O}(N_{\rm C}^2 N_{\rm V}^4)$ for the common cases with $N_{\rm A} < N_{\rm C} < N_{\rm V}$,\cite{Li:2016hb} where $N_{\rm C}$, $N_{\rm A}$, and $N_{\rm V}$ stand for the number of core, active, and virtual orbitals, respectively.
If $N_{\rm V}$ is brought down to $N_{\rm V}' < N_{\rm V}$, the overall cost of commutator evaluations will be effectively reduced by a factor of $(N_{\rm V}'/N_{\rm V})^4$.

To systematically truncate the virtual orbital space, we generalize the FNO idea of single-reference methods\cite{Taube:2005hv,DePrinceIII:2013ki} to the multireference setting.
In particular, we determine the natural orbital (NO) basis using the unrelaxed 1-RDM of the second-order perturbation theory (PT2) of SA-DSRG.
This unrelaxed 1-RDM is defined as the partial derivative of the second-order scalar term of $\bar{H}(s)$ [$\bar{H}_0^{[2]} (s)$] with respect to the SA Fock matrix elements [$\tens{f}{pq}{(0)}$]:
\begin{align}
\label{eq:gamma}
\dadapt{[2]}{pq} (s) = \partial \bar{H}_0^{[2]} (s) / \partial \tens{f}{pq}{(0)}.
\end{align}
Here, indices $p$ and $q$ label the semicanonical MOs.\cite{Li:2018kl}
We note that only the virtual-virtual block of $\dadapt{[2]}{pq} (s)$ need to be constructed and the explicit working equations are derived in Appendix \ref{sec:appendix1}.
The cost of forming the second-order unrelaxed 1-RDM scales as ${\cal O}(N_{\rm C}^2 N_{\rm V}^3)$, which is negligible compared to that of Eq.~\eqref{eq:O2T2C2}.
Nonetheless, there is a ${\cal O}(N_{\rm A}^6)$ dependence in the exact $\Gamma^{[2]}_{pq} (s)$ expressions due to the SA three-body density cumulant of the reference ensemble ($\boldsymbol \Lambda_3$).
The computation of $\boldsymbol \Lambda_3$ poses a potential bottleneck for applications with large active spaces, e.g., $N_{\rm A} > 40$.
Although treating such cases is beyond the scope of this work, we shall ignore the $\boldsymbol \Lambda_3$ contributions to $\dadapt{[2]}{pq} (s)$ only when the same procedure (or approximation) is applied to the final SA-DSRG energy.
For instance, no $\boldsymbol \Lambda_3$ is necessary in a conventional SA-DSRG computation on vertical transition energies.\cite{Wang2023}
In this situation, we also skip the $\boldsymbol \Lambda_3$ terms in $\dadapt{[2]}{pq} (s)$ for FNO SA-DSRG to avoid the additional cost.
Ignoring $\boldsymbol \Lambda_3$ may violate the positive definiteness of $\dadapt{[2]}{pq} (s)$.
However, we find that the magnitude of the negative eigenvalues of $\dadapt{[2]}{pq} (s)$ is of the order of $10^{-6}$, ignoring which leaves the final SA-DSRG energies largely unaffected.

The SA-DSRG-PT2 natural orbitals are obtained by solving the eigenvalue equation
\begin{align}
\label{eq:no}
{\boldsymbol \Gamma^{[2]}} (s) {\bf V}(s) = {\bf n}(s) {\bf V}(s),
\end{align}
where ${\bf n}(s)$ is the diagonal matrix of natural occupation numbers and the orthogonal matrix ${\bf V}(s)$ rotates the semicanonical MOs to the NO basis.
We then eliminate the columns of ${\bf V}(s)$ that correspond to occupation numbers (ONs) smaller than the user-defined cutoff value $n_\kappa$.
Subsequently, $\bar{H}(s)$ is built in this reduced set of virtual orbitals.
For convenience, these orbitals are recanonicalized to make the virtual-virtual block of the truncated SA Fock matrix diagonal.
Other than directly specifying the natural ON cutoff, $n_\kappa$ can be implicitly derived by the following metrics.\cite{Landau2010,Segarra-Marti2015}
Given that there are $N_{\rm V}'$ NOs with ONs greater than $n_\kappa$, the percentage of retained virtual orbitals is calculated as
\begin{align}
\label{eq:pv}
p_{\rm v} = (N_{\rm V}' / N_{\rm V}) \times 100\%.
\end{align}
Alternatively, we may measure the percentage of cumulative ONs using
\begin{align}
\label{eq:po}
p_{\rm o} = (\sum_{p}^{N_{\rm V}'} n_p / \sum_{p}^{N_{\rm V}} n_p) \times 100\%.
\end{align}
In practice, specifying a hard $p_{\rm o}$ or $p_{\rm v}$ threshold may break the orbital degeneracy for systems with high point group symmetry.
To this end, we slightly expand the retained virtual space by including all near-degenerate orbitals when the difference between successive ONs is less than 0.01 times the larger one.

The last piece of our FNO formulation is to estimate the truncation error introduced in the DSRG transformed Hamiltonian.
A well-educated guess comes from the SA-DSRG-PT2 results, obtained by subtracting the second-order Hamiltonian in the truncated FNO basis ($\bar{H}_{\rm PT2}^{\rm FNO}$) from that in the full basis ($\bar{H}_{\rm PT2}^{\rm MO}$).
This step introduces additional computational cost, yet it only scales as ${\cal O} [(N_{\rm V}^2 (N_{\rm C}^2 + N_{\rm A}^4)]$ because nothing but those components labeled by active indices are required to solve Eq.~\eqref{eq:Hbar_relax}.\cite{Li:2017bx}
Recent numerical benchmarks suggest that different flow parameters should be adopted for high accuracy in SA-DSRG-PT2 and SA-LDSRG(2) computations.\cite{Wang2023} 
Therefore, we adopt two flow parameters in this work:
one ($s_1$) for those related to SA-DSRG-PT2 and the other ($s_2$) for high-level DSRG methods in the truncated FNO basis (implicitly dependent on $s_1$).
Here, the high-level (HL) theory can be either LDSRG(2) or the third-order perturbation theory (PT3) and the final Hamiltonian reads as
\begin{align}
\label{eq:fno_corrected}
\bar{H}^{\rm MO}_{\rm HL}(s_2) \approx \bar{H}^{\rm FNO}_{\rm HL}(s_2,s_1) + [ \bar{H}^{\rm MO}_{\rm PT2}(s_1) - \bar{H}^{\rm FNO}_{\rm PT2}(s_1) ].
\end{align}
In this expression, all active-space DSRG Hamiltonians are represented in the same basis, chosen for convenience to be the one obtained from the initial SA-CASSCF computation.

\section{Numerical Results}
\label{sec:results}

The SA-DSRG-PT2 unrelaxed 1-RDM and the proposed FNO procedure were implemented in the open-source code \forte,\cite{FORTE2022} where the one- and two-electron integrals were obtained from the quantum-chemistry package \PSI.\cite{Smith:2020ci}
Throughout this work, we assumed density-fitted (DF) two-electron integrals for both SA-CASSCF and SA-DSRG computations using a common auxiliary basis set.
Core orbitals were excluded from the electron correlation treatment of DSRG according to the design of the corresponding basis set.
In the following, we mainly consider the sequential variant of LDSRG(2) [sq-LDSRG(2)], where a stepwise unitary transformation was performed $\bar{H}(s) = e^{-\hat{A}_2(s)} [e^{-\hat{A}_1(s)} \hat{H} e^{\hat{A}_1(s)}] e^{\hat{A}_2(s)}$.
Unless otherwise stated, we invoked the non-interacting virtual approximation to further speed up the sq-LDSRG(2) computations, where intermediate two-body operators [Eq.~\eqref{eq:O2T2C2}] labeled by three and four virtual indices were ignored.\cite{Zhang:2019wj}

\subsection{Vertical transitions of small organic molecules}
\label{subsec:questdb}

We first estimate the reliable choice of the FNO truncation threshold and the associated flow parameter $s_1$ [Eq.~\eqref{eq:fno_corrected}] by computing the first $\rm ^1 B_u$ excited state of 1,3-butadiene using SA-DSRG-PT3.
In particular, we consider $p_{\rm v}$ values ranging from 10 to 95\% (in increments of 2--5\%), while the $s_1$ parameter was scanned in 0.1~\sunit increments in the range of $[0.2, 2.0]$~\sunit.
The second flow parameter used in SA-DSRG-PT3 was fixed at $s_2 = 2.0$ \sunit, which is shown to yield accurate vertical transition energies (VTEs) in a recent benchmark study.\cite{Wang2023}
Three $\rm ^1 A_g$ and two $\rm ^1 B_u$ states were averaged for both SA-CASSCF and SA-DSRG-PT3.
Dunning's cc-pV$X$Z ($X =$ T, Q, 5) basis sets\cite{Dunning:1989bx} was adopted along with the respective cc-pV$X$Z-JKFIT auxiliary basis sets.\cite{Weigend:2002ga}
The geometry of 1,3-butadiene was taken from Ref.~\citenum{Loos2020}.

\begin{figure*}[ht!]
\centering
\includegraphics[width=0.75\textwidth]{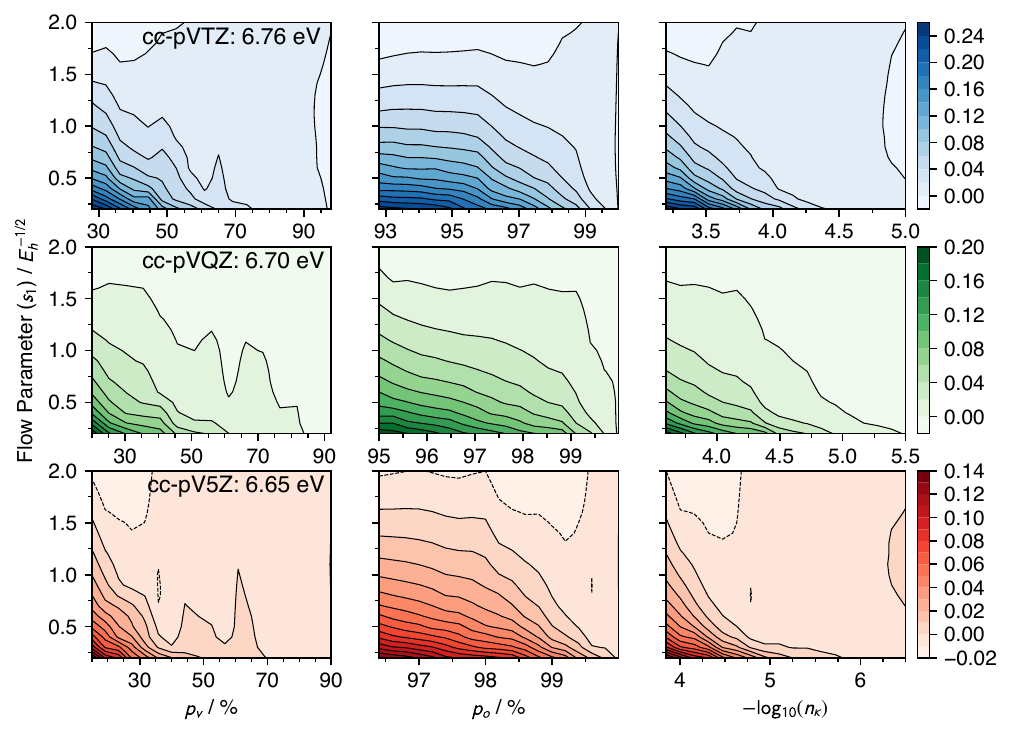}
\caption{The VTE error (in eV) of the lowest $\rm ^1 B_u$ state computed using the FNO truncated SA-DSRG-PT3 method against that of the full virtual space as a function of various FNO thresholds and flow parameters $s_1$. The other flow parameter $s_2$ is fixed at 2.0 \sunit. The cc-pVTZ, cc-pVQZ, and cc-pV5Z results are color coded in blue, green, and red, respectively. The VTEs of the complete SA-DSRG-PT3 with different basis sets are shown in the left column.}
\label{fig:butadiene_1bu}
\end{figure*}

Figure \ref{fig:butadiene_1bu} depicts the error of FNO SA-DSRG-PT3 VTEs for the lowest $\rm ^1 B_u$ state of butadiene as a function of $p_{\rm v}$ [Eq.~\eqref{eq:pv}] and $s_1$ [Eq.~\eqref{eq:fno_corrected}].
As expected, for all values of $s_1$ the errors generally decrease as more virtual orbitals are treated as unfrozen.
Interestingly, for a given VTE error we observe a negative correlation between $p_{\rm v}$ and $s_1$.
It is then preferable to construct the SA-DSRG-PT2 NOs using first-order amplitudes that are reasonably converged with respect to $s_1$ so that the virtual space can be significantly reduced with little penalty to the accuracy.
For example, the choice of $s_1 \sim 1.5$~\sunit, a value larger than the one used in typical SA-DSRG-PT2 computations,\cite{Li:2015iz,Wang2023} yields absolute errors within 0.02~eV for all $p_{\rm v}$ values presented in Figure~\ref{fig:butadiene_1bu}.
As pointed out in the previous work,\cite{Landau2010} using an identical $p_{\rm v}$ threshold leads to inconsistent energy errors with respect to the size of the basis set.
To this end, the system-specific $p_{\rm o}$ criterion [Eq.~\eqref{eq:po}] provides well-controlled VTE errors among various basis sets,\cite{Landau2010,Huntington:2015fq} as shown in the middle column of Figure~\ref{fig:butadiene_1bu}.
When $p_{\rm o} \in [97, 99.5]$\%, we calculate the energy deviation between any two basis sets to be $\leq 0.03$~eV.
As the size of the basis set increases, more orbitals tend to possess minuscule natural ONs (see the right column of Figure~\ref{fig:butadiene_1bu}), making the truncation scheme based on cumulative ONs more effective.
For example, $p_{\rm o} = 99$\% corresponds to 57, 38, and 25\% of the virtual orbitals of the cc-pVTZ, cc-pVQZ, and cc-pV5Z basis sets, respectively, which translates to a theoretical cost reduction of evaluating the contraction in Eq.~\eqref{eq:O2T2C2} by a factor of 9.5, 48, and 256, respectively.
In the following, we shall mainly adopt the $p_{\rm o}$ metric to characterize the virtual space truncation.

We further benchmarked the FNO-based SA-DSRG methods on 280 VTEs of various small organic molecules included in the QUEST database (QUESTDB).\cite{Loos2019,Loos2020,Veril2021,Sarkar2021a,Boggio-Pasqua2022}
As the previous test on 1,3-butadiene suggested, we focused on the $p_{\rm o}$ and $s_1$ values in the range of 97.0--99.5\% and 1.0--1.75~\sunit, respectively.
The flow parameters of the high-level SA-DSRG-PT3 and SA-sq-LDSRG(2) approaches were respectively set to 2.0 and 1.5 \sunit.\cite{Wang2023}
Consistent with QUESTDB, the aug-cc-pVTZ basis set\cite{Kendall:1992bj} was employed, except that we took advantage of the DF implementation of DSRG employing the aug-cc-pVTZ-JKFIT auxiliary basis set.
Molecular geometries, active spaces, and the number of averaged states were directly taken from QUESTDB\cite{Loos2019,Loos2020,Sarkar2021a} and summarized in the Supporting Information of Ref.~\citenum{Wang2023}.

\begin{figure}[ht!]
\centering
\ifpreprint
\includegraphics[width=0.5\columnwidth]{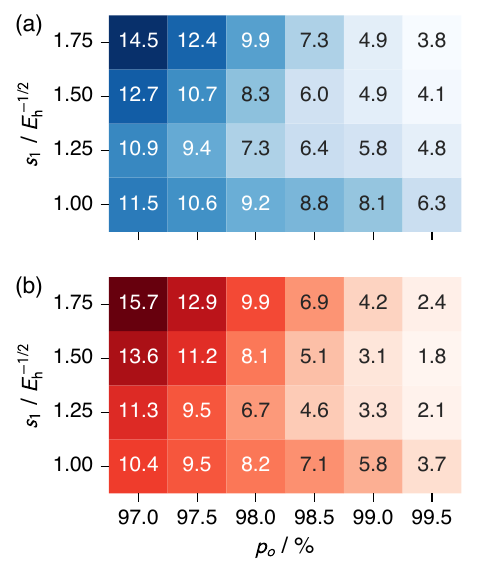}
\else
\includegraphics[width=0.8\columnwidth]{loos_mae}
\fi
\caption{The mean absolute error (in meV) of 280 VTEs in QUESTDB for the FNO-based (a) SA-DSRG-PT3 ($s_2 = 2.0$) and (b) SA-sq-LDSRG(2) ($s_2 = 1.5$)  relative to the corresponding untruncated counterparts.}
\label{fig:loos_mae}
\end{figure}

\begin{figure}[ht!]
\centering
\ifpreprint
\includegraphics[width=0.5\columnwidth]{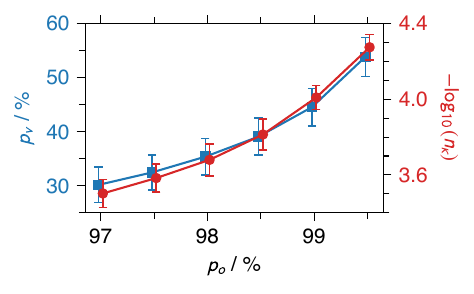}
\else
\includegraphics[width=\columnwidth]{loos_fno}
\fi
\caption{Comparison between various FNO truncation strategies. Data points were averaged over 238 unique computations to obtain the 280 VTEs of QUESTDB. The flow parameter $s_1$ was set to 1.5 \sunit. The error bars suggest the standard deviations of the data.}
\label{fig:loos_fno}
\end{figure}

Figure~\ref{fig:loos_mae} reports the mean absolute error (MAE) of deviations of all 280 VTEs for the FNO SA-DSRG methods from the respective untruncated analogs with varying $p_{\rm o}$ and $s_1$ values.
For all ($p_{\rm o}$, $s_1$) combinations, the FNO procedure yields highly accurate results with an MAE less than 0.02~eV.
The VTEs of FNO SA-DSRG become essentially indistinguishable ($\text{MAE} < 0.01$~eV) from those of the conventional SA-DSRG counterparts for $p_{\rm o} \geq 98.0\%$, where only 35\% of the virtual NOs are required (see Figure~\ref{fig:loos_fno}).
Interestingly, we observe an almost linear dependence between $p_{\rm v}$ and $-\log_{10}(n_\kappa)$ in Figure~\ref{fig:loos_fno}.
We also see no significant differences ($< 0.01$~eV) between valence and Rydberg states on the error statistics due to virtual space truncation (see Supporting Information).
Thus, the FNO implementation leaves the accuracy of SA-DSRG-PT3 and SA-sq-LDSRG(2) largely unaffected, both of which yield a 0.10~eV MAE on VTEs against the theoretical best estimates.\cite{Wang2023}.

\begin{figure}[ht!]
\centering
\ifpreprint
\includegraphics[width=0.5\columnwidth]{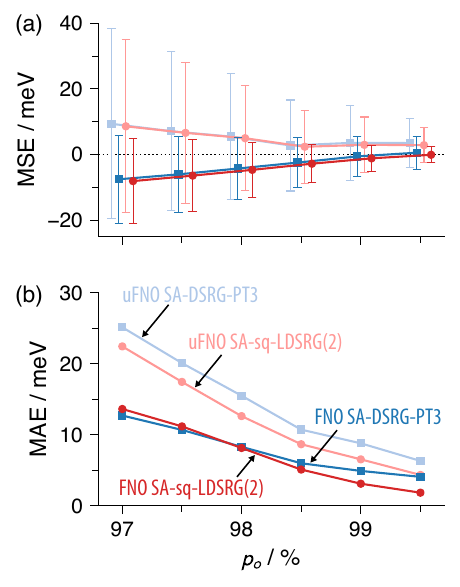}
\else
\includegraphics[width=0.9\columnwidth]{loos_depo}
\fi
\caption{The (a) mean signed error (MSE) and (b) mean absolute error (MAE) of 280 VTEs in QUESTDB for the FNO SA-DSRG variants relative to the corresponding untruncated counterparts. The error bars in the upper panel indicate the standard deviations of errors. The flow parameter $s_1$ was fixed at 1.5 \sunit.}
\label{fig:loos_depo}
\end{figure}

Figure~\ref{fig:loos_depo} presents the FNO SA-DSRG results without the second-order correction [i.e., $\bar{H}^{\rm MO}_{\rm HL}(s_2,s_1) \approx \bar{H}^{\rm NO}_{\rm HL}(s_2,s_1)$], labeled as uFNO.
In general, the second-order correction roughly cuts the standard deviation of errors in half and significantly reduces the MAE, especially for $p_{\rm o} < 98$\%.
Overall, we recommend choosing $p_{\rm o} \in [98, 99]$\% and $s_1 \in [1.25, 1.50]$~\sunit for a balance between accuracy and efficiency.
Unless otherwise mentioned, the flow parameter used to generate the SA-DSRG-PT2 NOs was set to 1.5 \sunit in the remaining computations of this work.

\subsection{Vertical transitions of transition-metal complexes}
\label{subsec:tm}
\begin{figure}[ht!]
\centering
\ifpreprint
\includegraphics[width=0.5\columnwidth]{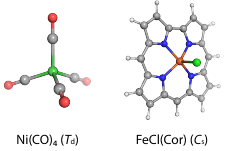}
\else
\includegraphics[width=0.85\columnwidth]{tmc}
\fi
\caption{Transition-metal complexes studied in this work.}
\label{fig:tmc}
\end{figure}

Transition-metal complexes provide rich photophysical and photochemical properties that can be utilized for a wide range of applications, such as energy conversion,\cite{Wenger2018} photocatalysis,\cite{Twilton2017} bioimaging,\cite{Xue2022} and phototherapy.\cite{Imberti2020}
The close interplay between metal centers and ligands results in densely populated electronic states of distinct nature, including metal-centered states, intra-ligand states, and metal-ligand/ligand-ligand charge transfer states.\cite{Mai2018a}
High-level quantum chemical methods are generally required to correctly predict their electronic structure and photochemical behaviors.
Here, we focus on the low-lying singlet states of nickel tetracarbonyl [\ce{Ni(CO)4}] and the spin splittings of chloroiron corrole [FeCl(Cor)] (see Figure~\ref{fig:tmc}).

\subsubsection{Nickel tetracarbonyl}

Nickel tetracarbonyl is well-known to photodissociate under excitation by near-ultraviolet (UV) light.
Despite significant efforts,\cite{Fuss2001,Cole-Filipiak2022,Hada1995,McKinlay2015} the excited-state mechanism of this process is still not fully understood.
One of the challenges is to characterize every transition in the experimental UV absorption spectrum, which generally lacks spectral resolution.\cite{Schreiner1968,Lever1979,Kotzian1989}
Three peaks can be identified in the gas-phase experiment:\cite{Kotzian1989} a principal one at 6.0~eV and two shoulders located at 5.4 and 4.6~eV.
All three bands are assigned to electron transfer excitations from nickel 3d to ligand $\pi^\ast$ orbitals.\cite{Kotzian1989,Hada1995,Pierloot1996}
A variety of theoretical methods have been employed to study the excited state of \ce{Ni(CO)4},\cite{Hada1995,Pierloot1996,VanGisbergen1999,Nooijen2000,McKinlay2015}
yet accurately reproducing the experimental spectrum appears arduous.
For example, LR-CCSD suggests that the first bright state lies 5.05~eV above the ground state,\cite{McKinlay2015} 0.46~eV higher than the experimental value.\cite{Kotzian1989}
Surprisingly, the inclusion of perturbative triples increases the LR-CCSD value by 1.26~eV, further enlarging the discrepancy between theory and experiment.\cite{McKinlay2015}
Nonetheless, all computations reveal a high density of states within 2~eV above the first excited state.

In the following, we computed a number of low-lying singlet states of \ce{Ni(CO)4} using the FNO SA-DSRG methods.
We adopted a CAS(10e,13o) reference space, which is consistent with previous work of complete-active-space second-order perturbation theory (CASPT2).\cite{Pierloot1996}
The active orbitals include five Ni 3d orbitals and eight $\pi^\ast$ orbitals of CO, five of which are strongly mixed with Ni 4d orbitals (see Supporting Information).
Forty states were averaged in the SA-CASSCF and SA-DSRG computations without point group symmetry.
Scalar relativistic effects were addressed using the second-order Douglas--Kroll--Hess (DKH2) Hamiltonian.\cite{Wolf:2002ia}
The main basis set was constructed from the cc-pwCVQZ-DK basis set\cite{Balabanov:2005hm} for Ni and the cc-pVQZ-DK basis set\cite{deJong:2001gp} for C and O atoms.
Because we mainly focus on valence transitions from nickel 3d to ligand $\pi^\ast$ orbitals, diffuse functions were found to have little impact on the vertical transition energy (see Supporting Information).
Thirteen orbitals (Ni 1s2s2p, C 1s, and O 1s) were excluded from electron-correlation treatments.
To enable DF, an auxiliary basis set was generated by the AutoAux procedure,\cite{Stoychev2017} which was available in the Python library of Basis Set Exchange.\cite{Pritchard:2019gs}
The bond lengths were set to experimental values:\cite{Hedberg1979} 1.838~\r{A} for \ce{Ni-C} and 1.141~\r{A} for \ce{C-O}, as done in previous computations.\cite{Hada1995,Pierloot1996,Nooijen2000}

\begin{table}[t!]
\begin{threeparttable}
\centering
\ifpreprint
\small
\renewcommand{\arraystretch}{1.0}
\else
\scriptsize
\renewcommand{\arraystretch}{1.25}
\fi
\caption{Error statistics (in eV) for the 16 vertical excitation energies of \ce{Ni(CO)4} obtained using FNO-based SA-sq-LDSRG(2) ($s_2 = 1.5$~\sunit) deviated from those without virtual orbital space truncation.}
\label{table:Ni_FNO}
\begin{tabular*}{\columnwidth}{@{\extracolsep{\stretch{1}}} l *{4}{d{2.2}} d{1.2} @{}}
\hline
\hline
& \multicolumn{5}{c}{$p_{\rm o}$ / \%} \\
\cline{2-6}
Metric & \multicolumn{1}{c}{96} & \multicolumn{1}{c}{97} & \multicolumn{1}{c}{98} & \multicolumn{1}{c}{99} & \multicolumn{1}{c}{100} \\
\hline
MAE\tnote{a} & 0.09 & 0.06 & 0.05 & 0.02 \\
MIN\tnote{a} & -0.10 & -0.07 & -0.05 & -0.03 \\
MAX\tnote{a} & -0.08 & -0.06 & -0.04 & -0.02 \\
\hline
$\log_{10}(n_\kappa)$ & -3.38 & -3.49 & -3.72 & -4.11 \\
$N_{\rm V}'$\tnote{b} & \multicolumn{1}{r}{155} & \multicolumn{1}{r}{178} & \multicolumn{1}{r}{210} & \multicolumn{1}{r}{275} & \multicolumn{1}{r}{539} \\
Time\tnote{c} & \multicolumn{1}{r}{8} & \multicolumn{1}{r}{7} & \multicolumn{1}{r}{14} & \multicolumn{1}{r}{23} & \multicolumn{1}{r}{127} \\
\hline
\hline
\end{tabular*}
\begin{tablenotes}
\item [a] $\text{MAE} = \frac{1}{16} \sum_{i=1}^{16} |\Delta_i|$, $\text{MIN} = \min(\Delta_i)$, and $\text{MAX} = \max(\Delta_i)$ with $\Delta_i = x_i (p_{\rm o}) - x_i (100)$.
\item [b] Number of retained virtual orbitals after FNO truncation.
\item [c] Minutes per iteration of amplitudes update using 48 cores and 256~GB of memory on a node with dual AMD EPYC 7H12 and 512~GB of memory.
\end{tablenotes}
\end{threeparttable}
\end{table}

Table \ref{table:Ni_FNO} reports the FNO-SA-sq-LDSRG(2) results for various FNO threshold values.
For a given $p_{\rm o}$ value, all the FNO-SA-sq-LDSRG(2) VTEs are consistently red-shifted by almost the same amount (within less than 0.02~eV).
Importantly, the state ordering is preserved despite the FNO cutoff.
As expected, the VTE error decreases as the virtual orbital space approaches the full one and the MAE becomes smaller than 0.05~eV for $p_{\rm o} > 98\%$.
As found in the benchmarks on QUESTDB, the FNO formalism is most effective at eliminating the least occupied orbitals whose ONs amount to 1--2\% of the total virtual ONs.
For $p_{\rm o} < 98\%$, the decrease in the number of virtual orbitals slows down while noticeable absolute errors to VTEs emerge.

\begingroup
\begin{table*}[t!]
\begin{threeparttable}
\centering
\ifpreprint
\scriptsize
\renewcommand{\arraystretch}{1.0}
\else
\scriptsize
\renewcommand{\arraystretch}{1.25}
\fi

\caption{Vertical transition energies (in eV) and oscillator strengths (in parentheses) of \ce{Ni(CO)4} computed using various theoretical methods. The CAS(10e,13o) active space is used for all MR methods.}
\label{table:Ni_compare}

\begin{tabular*}{\textwidth}{@{\extracolsep{\stretch{1}}} l d{2.4} d{2.4} d{2.3} d{2.4} d{2.4} d{2.4} d{2.4} d{1.2} @{}}
\hline
\hline

& & & & \multicolumn{4}{c}{SA-DSRG\tnote{c,e}} & \\
\cline{5-8}
State\tnote{a} & \multicolumn{1}{c}{LR-CCSD\tnote{b}} & \multicolumn{1}{c}{CASSCF\tnote{c}} & \multicolumn{1}{c}{CASPT2\tnote{d}} & \multicolumn{1}{c}{PT2} & \multicolumn{1}{c}{PT3} & \multicolumn{1}{c}{sq-L(2)} & \multicolumn{1}{c}{sq-rL(2)} & \multicolumn{1}{c}{Exp.\tnote{f}} \\
\hline

$\rm ^1 E (t_2 \rightarrow t_2)$ & 4.77 & 6.37 & 5.03 & 5.32 & 5.04 & 5.30 & 5.13 & \\
$\rm ^1 T_1 (t_2 \rightarrow t_2)$ & 4.73 & 6.47 & 4.85 & 5.32 & 5.06 & 5.31 & 5.15 & \\
$\rm ^1 T_1 (t_2 \rightarrow e)$ & 5.10 & 6.74 & 5.20 & 5.64 & 5.31 & 5.59 & 5.38 &  \\
\multirow{2}{*}{$\rm ^1 T_2 (t_2 \rightarrow t_2)$}
 & 5.05 & 6.85 & 5.34 & 5.62 & 5.36 & 5.62 & 5.44 & 4.6 \\
 & (0.045) & (0.072) & (0.03) & (0.085) & (0.062) & (0.067) & (0.036) & \\
$\rm ^1 T_1 (e \rightarrow t_2)$ & 5.43 & 6.62 & 5.61 & 5.79 & 5.41 & 5.78 & 5.69 & \\
$\rm ^1 A_1 (e \rightarrow e, t_2 \rightarrow t_2)$ & 5.64 & 7.09 & 5.53 & 5.92 & 5.60 & 5.94 & 5.75 & \\
\multirow{2}{*}{$\rm ^1 T_2 (t_2 \rightarrow e)$}
 & 5.53 & 7.21 & 5.61 & 6.01 & 5.65 & 5.97 & 5.76 & 4.6? \\
 & (0.095) & (0.004) & (0.16) & (0.018) & (0.021) & (0.022) & (0.041) & \\
$\rm ^1 A_2 (t_2 \rightarrow t_1, e \rightarrow e)$ & 5.98 & 6.94 & 6.30 & 6.28 & 5.98 & 6.24 & 6.13 & \\
\multirow{2}{*}{$\rm ^1 T_2 (e \rightarrow t_2)$}
 & 5.93 & 7.46 & 5.75 & 6.38 & 5.97 & 6.35 & 6.17 & 4.6? \\
 & (0.082) & (0.054) & (0.15) & (0.019) & (0.043) & (0.016) & (0.028) & \\
$\rm ^1 T_1 (t_2 \rightarrow t_1)$ & 6.27 & 7.13 & 5.77 & 6.38 & 6.19 & 6.36 & 6.27 & \\
$\rm ^1 E (t_2 \rightarrow t_1)$ & 6.21 & 7.16 & 6.14 & 6.42 & 6.18 & 6.40 & 6.30 & \\
$\rm ^1 A_2 (e \rightarrow e, t_2 \rightarrow t_1)$ & 6.51 & 7.90 & 5.90 & 6.73 & 6.34 & 6.68 & 6.49 & \\
\multirow{2}{*}{$\rm ^1 T_2 (t_2 \rightarrow t_1)$}
 & 6.58 & 7.59 & 6.19 & 6.76 & 6.49 & 6.74 & 6.64 & 5.4 \\
 & (0.228) & (0.075) & (0.20) & (0.237) & (0.180) & (0.225) & (0.236) & \\
$\rm ^1 E (e \rightarrow e)$ & 6.62 & 8.27 & 5.99 & 6.93 & 6.45 & 6.88 & 6.64 & \\
$\rm ^1 T_1 (e \rightarrow t_1)$ & 7.37 & 7.80 & 7.03 & 7.21 & 6.80 & 7.16 & 7.12 & \\
\multirow{2}{*}{$\rm ^1 T_2 (e \rightarrow t_1)$}
 & 7.50 & 8.38 & 6.40 & 7.45 & 7.03 & 7.40 & 7.32 & 6.0 \\
 & (0.175) & (0.902) & (2.01) & (0.393) & (0.392) & (0.402) & (0.351) & \\

\hline
\hline
\end{tabular*}
\begin{tablenotes}
\item [a] The order and excitation characters are sorted based on the SA-sq-LDSRG(2) ($s = 1.5$) results.
\item [b] Computed using the latest version of the Dalton package.\cite{Aidas2014} The cc-pwCVTZ-DK and cc-pVTZ basis sets were employed for Ni atom and other atoms, respectively. Scalar relativistic effects were treated via the DKH2 Hamiltonian. Two Rydberg states are observed: $\rm ^1 T_2 (t_2 \rightarrow a_1)$ 6.19~eV (0.022) and $\rm ^1 E (e \rightarrow a_1)$ 7.24~eV.
\item [c] Averaged over 40 singlet states.
\item [d] Taken from Ref.~\citenum{Roos:1996ea}. Computations were performed using the ANO-S type basis sets and the orbitals from SA-CASSCF averaged within every irreducible representation of $D_2$.\cite{Roos:1996ea} The CASPT2 level shift was set to 0.3~\Eh. Oscillator strengths were obtained via CAS state interaction.
\item [e] Abbreviations: PT2 = SA-DSRG-PT2 ($s = 0.5$), PT3 = SA-DSRG-PT3~($s = 2.0$), sq-L(2) = SA-sq-LDSRG(2)~($s = 1.5$), sq-rL(2) = SA-sq-LDSRG(2)~($s = 3.5$, relaxed).
\item [f] Ref.~\citenum{Kotzian1989}.
\end{tablenotes}
\end{threeparttable}
\end{table*}
\endgroup

In Table \ref{table:Ni_compare}, we report the VTEs of \ce{Ni(CO)4} computed using various SA-DSRG methods without the FNO approximation.
We take the SA-sq-LDSRG(2) (relaxed, $s = 3.5$~\sunit) data as the best values predicted by SA-DSRG and the reason for this choice is discussed in the Supporting Information.
In general, SA-sq-LDSRG(2) ($s=1.5$~\sunit) overestimates the VTEs by 0.1--0.2~eV because energies are not converged with respect to the flow parameter.
With the optimal flow parameters benchmarked on QUESTDB,\cite{Wang2023} the SA-DSRG-PT3 results are underestimated by 0.16~eV on average, while excellent agreement is found between the SA-DSRG-PT2 and SA-sq-LDSRG(2) values, with a max deviation of 0.06~eV.
For CASPT2, most VTEs are significantly lower than the SA-DSRG-PT2 predictions, possibly caused by the state-specific procedure used in CASPT2.
Similar discrepancies ($\sim$0.3~eV) are observed between LR-CCSD and SA-DSRG for those states with VTEs smaller than 6.0~eV.
In LR-CCSD, states that arise from $\rm e/t_2 \rightarrow t_2$ transitions possess a reasonable amount of Rydberg character, which might not be easily recovered in the SA-DSRG treatment without including an additional set of $\pi^\ast$ orbitals in the active space.

We now focus on the states of $\rm ^1 T_2$ symmetry as it is the only dipole-allowed transition from the ground state ($\rm ^1 A_1$).
The best SA-DSRG approach predicts the first $\rm ^1 T_2$ state lying 5.44~eV above the ground state, 0.8~eV higher than the gas-phase experimental value.
The LR-CCSD method overestimates the experimental value by roughly 0.5~eV.
Previous CASPT2 studies showed that the use of state-specific CASSCF orbitals reduces the VTEs for $\rm 3d \rightarrow t_2/e$ states by as large as 0.9~eV over those based on state-averaged orbitals.\cite{Roos:1996ea,Pierloot1996}
The large energy shift reflects the importance of excited-state orbital relaxation effects for \ce{Ni(CO)4}.
It is plausible that similar energy shifts may also be observed for SA-DSRG.

\begin{figure}[h!]
\centering
\ifpreprint
\includegraphics[width=0.5\columnwidth]{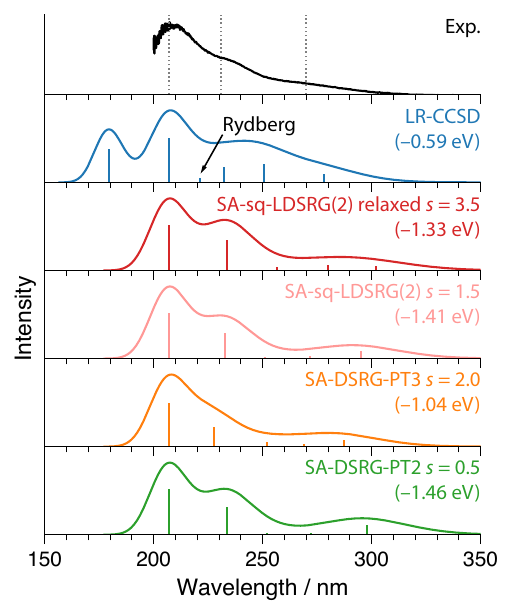}
\else
\includegraphics[width=0.9\columnwidth]{ni_spec}
\fi
\caption{Experimental and theoretical absorption spectra of \ce{Ni(CO)4}. The theoretical spectra are convoluted with Gaussians of 0.6~eV full width at half maximum. Every theoretical spectrum is shifted by the value given in parentheses to align the $\rm ^1 T_2$ state of the largest oscillator strength with the experimental main peak (207~nm). The experimental spectrum is adapted with permission from \textit{J.~Am.~Chem.~Soc.}~\textbf{1989}, \textit{111}, 7687--7696.\cite{Kotzian1989} Copyright 1989 Americal Chemical Society.}
\label{fig:ni_spec}
\end{figure}

The experimental spectrum is well reproduced by all SA-DSRG methods (see Figure~\ref{fig:ni_spec}).
Based on the SA-DSRG results, we believe the main peak at 207~nm is due to the $\rm e \rightarrow t_1$ excitation.
The first shoulder at 230~nm is dominated by $\rm t_2 \rightarrow t_1$ transitions, likely with a $\rm t_2 \rightarrow a_1$ Rydberg character (See Supporting Information).
The second shoulder (270~nm) consists of the lowest three $\rm ^1 T_2$ states, leading to a very wide band shape.

\subsubsection{Chloroiron corrole}

Our final example is chloroiron corrole [FeCl(Cor)].
Both experiments and theories reveal a triplet ground state from a d$^5$ $S = 3/2$ Fe(III) antiferromagnetically coupled with the (corrolate)$^{2-}$ radical.\cite{Zakharieva2002,Walker2006,Ye2008,Roos2008}
A close-lying quintet state is also suggested by CASPT2 with a spin gap of 0.33~eV.\cite{Roos2008}
Here, we computed several low-lying states of FeCl(Cor) using FNO SA-DSRG-PT3 with a basis set customized by mixing the ANO-RCC-VQZP basis set for Fe, Cl, and N atoms with the ANO-RCC-VTZP basis set for C and H.\cite{Roos2005}
Like the previous computations on \ce{Ni(CO)4}, we adopted the DKH2 Hamiltonian for scalar relativistic effects and the AutoAux method for generating auxiliary basis sets.
A CAS(12e,15o) active space was employed, which includes a full set of Fe 3d and 4d orbitals, one \ce{Fe-N} $\sigma^\ast$ orbital, one Cl 3p$_z$ orbital, and one $\pi$ and two $\pi^\ast$ orbitals of corrole.
Following Ref.~\citenum{Roos2008}, we use state-specific CASSCF orbitals (depicted in the Supporting Information).

As shown in Table \ref{table:FeCl}, the state ordering is reasonably consistent among post-MCSCF methods, especially for the lowest 7 states.
All CASPT2, DSRG-MRPT2 and FNO-DSRG-MRPT3 suggest a $\rm ^3 A''$ ground state and a $\rm ^5 A'$ state being the lowest excited state.
The triplet-quintet gap is predicted to be 0.3~eV by MRPT2s, 0.1~eV lower than that of FNO-DSRG-MRPT3.
The second quintet state $\rm 1\, ^5 A''$ appears near-degenerate to $\rm 1\, ^5 A'$, separated by less than 0.01~eV according to FNO-DSRG-MRPT3.
Overall, both DSRG methods find 6 low-energy excited states within 1.0~eV of the ground state.

\begin{table}[t!]
\begin{threeparttable}
\centering
\ifpreprint
\small
\renewcommand{\arraystretch}{1.0}
\else
\scriptsize
\renewcommand{\arraystretch}{1.25}
\fi

\caption{Vertical transition energies (in eV) of FeCl(Cor) computed using various theoretical methods.}
\label{table:FeCl}

\begin{tabular*}{\columnwidth}{@{\extracolsep{\stretch{1}}} l d{1.2} d{1.2} d{1.2} d{1.2} @{}}
\hline
\hline

& & & \multicolumn{2}{c}{DSRG} \\
\cline{4-5}
State & \multicolumn{1}{c}{CASSCF\tnote{a}} & \multicolumn{1}{c}{CASPT2\tnote{b}} & \multicolumn{1}{c}{PT2\tnote{a,c}} & \multicolumn{1}{c}{FNO PT3\tnote{a,d}} \\
\hline

$\rm 1\, ^3 A''$ & 0.00 & 0.00 & 0.00 & 0.00 \\
$\rm 1\, ^5 A'$ & -0.44 & 0.33 & 0.30 & 0.43 \\
$\rm 1\, ^5 A''$ & 0.32 & 0.54 & 0.40 & 0.43 \\
$\rm 1\, ^7 A'$ & -0.26 & 0.62 & 0.54 & 0.68 \\
$\rm 2\, ^3 A''$ & 0.74 &   & 0.78 & 0.70 \\
$\rm 1\, ^3 A'$ & 0.80 & 1.13 & 0.83 & 0.77 \\
$\rm 1\, ^7 A''$ & -0.03 & 1.23 & 0.87 & 0.90 \\
$\rm 2\, ^3 A'$ & 1.33 &   & 1.43 & 1.15 \\
$\rm 2\, ^5 A''$ & 1.05 & 1.35 & 1.18 & 1.15 \\
$\rm 1\, ^1 A''$ & 1.29 & 1.75 & 1.50 & 1.21 \\
$\rm 2\, ^5 A'$ & 1.09 & 1.66 & 1.22 & 1.21 \\
$\rm 1\, ^1 A'$ & 1.35 & 1.73 & 1.55 & 1.28 \\

\hline
\hline
\end{tabular*}
\begin{tablenotes}
\item [a] This work: CAS(12e,15o) active space with a mixed basis set of ANO-RCC-VQZP(Fe,Cl,N)/ANO-RCC-VTZP(C,H).
\item [b] Ref.~\citenum{Roos2008}: CAS(14e,14o) active space with a mixed basis set of ANO-RCC-VTZP(Fe,Cl,N)/ANO-RCC-VDZP(C,H).
\item [c] Flow parameter: $s = 0.5$~\sunit.
\item [d] FNO threshold: $p_{\rm o} = 99.0\%$, flow parameters: $s_1 = 1.5, s_2 = 0.5$~\sunit.
\end{tablenotes}
\end{threeparttable}
\end{table}

\section{Conclusions}
\label{sec:conclusion}

To summarize, we have introduced a low-cost frozen-natural-orbital implementation for the high-level SA-DSRG methods that go beyond second-order perturbation theory.
In the FNO approximation, we discard the virtual orbitals associated with small eigenvalues of the SA-DSRG-PT2 unrelaxed one-body reduced density matrix.
Benchmark computations of vertical transition energies of small organic molecules show that a threshold that retains 98--99\% of the total virtual occupation numbers yields a superior balance between accuracy and efficiency.
Specifically, the mean absolute error of VTE incurred by FNO is less than 0.01~eV using only 35--45\% of the total number of virtual orbitals, with an overall computational cost reduction of 4--9 times.
For the transition-metal complex \ce{Ni(CO)4}, the FNO scheme introduces an error of 0.02--0.05~eV to the VTEs with a speedup of 5--9 times the conventional approach.
Importantly, the relative state ordering is not affected by FNO for SA-DSRG.
The FNO approximation enables the SA-DSRG-PT3 and SA-LDSRG(2) methods to be easily applied on systems with more than 1000 basis functions, such as the \ce{FeN4} complexes.\cite{Phung2023}

We note however, the FNO approach generally yields discontinuities on the potential energy surface (PES).
In FNO-CASPT2, this problem is alleviated by using the threshold of cumulative occupation numbers.\cite{Segarra-Marti2015}
Preliminary tests on the FNO SA-LDSRG(2) potential energy curves of \ce{LiF} show sudden energy changes of $\sim$0.2~m\Eh on a grid of 0.01~\AA, which is caused by the inclusion of a single virtual orbital.
To this end, it may be beneficial to fix the number of virtual orbitals along the entire PES.

Besides the PES issue, the current FNO formalism can be optimized to further improve its efficiency and accuracy.
Given that the AutoAux procedure produces a large set of auxiliary basis functions, it is preferable to reduce its dimension using natural auxiliary functions (NAF).\cite{Kallay2014}
The NAF method is also expected to speed up the SA-DSRG-PT2 method for generating the 1-RDM and the DSRG Hamiltonian $\bar{H}(s)$.
Currently, we apply an additive second-order correction to $\bar{H}(s)$ to recover the correlation effect from the frozen virtual orbitals.
Recent work in FNO-CC suggests a superior multiplicative correction, reducing the FNO error by a factor of 3.\cite{Nagy2021}
It would be interesting to see if this approach performs equally well on correcting the FNO approximated DSRG Hamiltonian.
Finally, the natural occupation number from the SA-DSRG-PT2 1-RDM may be adopted as a criterion for selecting active orbitals in MR computations.\cite{Khedkar:2019ge}
The set of SA-DSRG-PT2 natural orbitals also provides an alternative to bypass the orbital optimization of MCSCF.\cite{Levine2021}

\section{Associated Content}

See supporting information for
1) VTEs of QUESTDB computed using various $(s_1, p_{\rm o})$ combinations,
2) CASSCF orbitals of \ce{Ni(CO)4} and FeCl(Cor).

\section{Acknowledgements}
C.L.\ and S.M.\ were supported by the National Natural Science Foundation of China (Grant No.~22103005) and the Fundamental Research Funds for the Central Universities.
R.H.\ and F.A.E. acknowledge support by the U.S. Department of Energy under Award No. DE-SC0019374.

\appendix


\setcounter{equation}{0}
\renewcommand{\theequation}{A\arabic{equation}}

\setcounter{table}{0}
\renewcommand{\thetable}{A\arabic{table}}

\section{Expressions for the second-order unrelaxed 1-RDMs}
\label{sec:appendix1}

We first introduce the state-averaged 1-RDM of the reference ensemble $\{\Psi_\alpha | \alpha = 1, 2, \dots, n\}$:
\begin{align}
\label{eq:sf1-RDM}
\dadapt{p}{q} = \sum_{\alpha}^{n} w_\alpha \braket{\Psi_\alpha | \squop{p}{q} | \Psi_\alpha},
\end{align}
where $w_\alpha$ is the weight of state $\Psi_\alpha$.
The spin-free single excitation operators are given by $\squop{p}{q} = \sum_{\sigma}^{\uparrow\downarrow} \sqop{p_\sigma}{q_\sigma}$ expressed in terms of creation ($\cop{p_\sigma}$) of annihilation ($\aop{q_\sigma}$) operators using the shorthand notation $\sqop{pq\cdots}{rs\cdots} = \cop{p}\cop{q}\cdots\aop{s}\aop{r}$ and the summation $\sigma$ runs over the two spin cases.
We then write out the normal-ordered second-quantized bare Hamiltonian
\begin{align}
\label{eq:Hbare}
\hat{H} = E_0 + \sum_{pq} \tens{f}{p}{q} \no{\squop{p}{q}}
+ \frac{1}{2} \sum_{pqrs} \braket{pq|rs} \no{\squop{pq}{rs}} ,
\end{align}
where $E_0 = \sum_{\alpha}^{n} w_\alpha \braket{\Psi_\alpha | \hat{H} | \Psi_\alpha}$ is the averaged reference energy and $\tens{f}{p}{q}$ are the SA Fock matrix elements defined by one- ($\tens{h}{p}{q}$) and two-electron ($\braket{pq|rs}$) integrals:
\begin{align}
\tens{f}{p}{q} = \tens{h}{p}{q} + \sum_{rs} \dadapt{r}{s} ( \braket{pr|qs} - \frac{1}{2} \braket{pr|sq}).
\end{align}
In Eq.~\eqref{eq:Hbare}, $\squop{pq}{rs} = \squop{p}{r} \squop{q}{s} - \kro{q}{r} \squop{p}{s}$ and operators embraced by curly brackets are normal-ordered based on the Mukherjee-Kutzelnigg formalism.\cite{Mukherjee:1997tk,Kutzelnigg:1997ut}

To derive the working equations of $\boldsymbol \Gamma^{[2]} (s)$ using Eq.~\eqref{eq:gamma}, we consider the second-order DSRG Hamiltonian:\cite{Li:2015iz,Li:2017bx}
\begin{align}
\label{eq:Hbar_2nd}
\bar{H}^{[2]} (s) = \hat{H} + [\hat{H}, \hat{A}^{(1)}(s)] + \frac{1}{2} [[\hat{H}^{(0)}, \hat{A}^{(1)}(s)], \hat{A}^{(1)}(s)].
\end{align}
The zeroth-order Hamiltonian has the form
\begin{align}
\hat{H}^{(0)} = E_0 + \sum_{mn}^{N_{\rm C}} \tens{f}{m}{n} \no{\squop{m}{n}} + \sum_{uv}^{N_{\rm A}} \tens{f}{u}{v} \no{\squop{u}{v}} + \sum_{ef}^{N_{\rm V}} \tens{f}{e}{f} \no{\squop{e}{f}}.
\end{align}
The first-order cluster amplitudes in $\hat{A}^{(1)}(s)$ are given by
\begin{align}
\label{eq:t_1_amps}
\tens{t}{a}{i,(1)} (s) &= \big[ \tens{f}{a}{i} + \frac{1}{2} \sum_{ux}^{N_{\rm A}} \tens{\Delta}{u}{x} \dadapt{x}{u} \tens{\check{t}}{ax}{iu,(1)} \big] \frac{1 - e^{-s(\tens{\Delta}{a}{i})^2}}{\tens{\Delta}{a}{i}}, \\
\label{eq:t_2_amps}
\tens{t}{ab}{ij,(1)} (s) &= \braket{ab|ij} \frac{ 1 - e^{-s(\tens{\Delta}{ab}{ij})^2} }{\tens{\Delta}{ab}{ij}},
\end{align}
where $\tens{\check{t}}{ab}{ij} = 2 \tens{t}{ab}{ij} - \tens{t}{ab}{ji}$ and the denominators $\tens{\Delta}{ab\cdots}{ij\cdots} = \dfock{i} + \dfock{j} + \cdots - \dfock{a} - \dfock{b} - \cdots$ are defined using semicanonical orbital energies $\dfock{p}$.
Indices $i$ and $j$ label either core or active orbitals, while indices $a$ and $b$ label either active or virtual orbitals.
Because $\hat{H}^{(0)}$ cannot fully contract with $\hat{A}^{(1)}(s)$ to generate nonzero scalar terms,\cite{Li:2015iz} the second-order corrections to the reference 1-RDM solely come from the doubly nested commutator in Eq.~\eqref{eq:Hbar_2nd}.
Namely, the diagonal blocks of $\boldsymbol \Gamma^{[2]} (s)$ are given by
\begin{align}
\dadapt{[2]}{pq} (s) = \dadapt{p}{q}
+ \underbrace{\frac{1}{2} [[\no{\squop{p}{q}}, \hat{A}^{(1)}(s)], \hat{A}^{(1)}(s)]_0}_{\dadapt{(2)}{pq} (s)} .
\end{align}
Specifically, the virtual-virtual block of $\dadapt{(2)}{pq} (s)$ can be expressed in terms of cluster amplitudes and SA density cumulants ($\cuadapt{pq\cdots}{rs\cdots}$) of the reference ensemble:
\begin{align}
\dadapt{(2)}{ef} =&\,
 2 \tens{t}{e}{m} \tens{t}{f}{m}
 + \tens{t}{e}{u} \tens{t}{f}{v} \dadapt{u}{v}
 + ( \tens{t}{e}{u} \tens{t}{fv}{xy} + \tens{t}{ey}{uv} \tens{t}{f}{x} ) \cuadapt{uv}{xy}
 + 2 \tens{t}{ee'}{mn} \tens{\check{t}}{fe'}{mn} \notag\\
 &
 + {\cal S} ( \tens{t}{ee'}{mu} \tens{\check{t}}{fe'}{mv} ) \dadapt{u}{v}
 + \tens{t}{ev}{mn} \tens{\check{t}}{fu}{mn} \cdadapt{u}{v}
+ \tens{t}{ee'}{uv} \tens{t}{fe'}{xy} \dadapt{uv}{xy}
\notag\\
&
+ \frac{1}{2} {\cal S} ( \tens{t}{ey}{mu} \tens{\check{t}}{fx}{mv} ) \dadapt{u}{v} \cdadapt{x}{y}
+ {\cal P} ( \tens{t}{ey}{mv} \tens{t}{fu}{mx} ) \cuadapt{uv}{xy}
\notag\\
&
+ \frac{1}{2} \tens{t}{ez}{uv} \tens{t}{fw}{xy} \cdadapt{w}{z} \cuadapt{uv}{xy}
+ \frac{1}{4} \tens{t}{ez}{ux} \tens{t}{fw}{vy} \dadapt{u}{v} \dadapt{x}{y} \cdadapt{w}{z}
\notag\\
&
+ \frac{1}{2} {\cal P} ( \tens{t}{ey}{wv} \tens{t}{fu}{zx} ) \dadapt{w}{z} \cuadapt{uv}{xy}
+ \tens{t}{ez}{uw} \tens{t}{fv}{xy} \cuadapt{xyz}{uvw} , \label{eq:Dvv}
\end{align}
where the summation over repeated indices is assumed and the superscript for perturbation order is omitted for brevity.
Core orbitals are labeled by indices $m, n$, active orbitals by indices $u, v, w, x, y, z$, and virtual orbitals by indices $e, f, e'$.
In Eq.~\eqref{eq:Dvv}, we introduce a few intermediates: $\cdadapt{u}{v} = 2\kro{u}{v} - \dadapt{u}{v}$, $\dadapt{uv}{xy} = \cuadapt{uv}{xy} + \dadapt{u}{x} \dadapt{v}{y} - \frac{1}{2} \dadapt{v}{x} \dadapt{u}{y}$, along with two short-hand notations ${\cal S} ( \tens{A}{pq}{rs} \tens{B}{ab}{cd} ) = \tens{A}{pq}{rs} \tens{B}{ab}{cd} + \tens{A}{pq}{sr} \tens{B}{ab}{dc}$ and ${\cal P} ( \tens{t}{au}{ij} \tens{t}{cd}{kv} ) = 2 \tens{t}{au}{ij} \tens{t}{cd}{kv} - \tens{t}{au}{ij} \tens{t}{cd}{vk} - \tens{t}{au}{ji} \tens{t}{cd}{kv} - \tens{t}{av}{ji} \tens{t}{cd}{uk}$.
Expressions for the two- and three-body SA density cumulants can be found in Ref.~\citenum{Li:2021sf}.
In the single-reference limit, Eq.~\eqref{eq:Dvv} reproduces the unrelaxed 1-RDM corrections due to the second-order M{\o}ller--Plesset perturbation theory.

\bibliography{york-bib,bib-ext}

\end{document}